# Magnetoelectric effect in layered ferrite/PZT composites. Study of the demagnetizing effect on the magnetoelectric behavior.

V. Loyau, V. Morin, G. Chaplier, M. LoBue, and F. Mazaleyrat

*SATIE UMR 8029 CNRS, ENS Cachan, Université Paris-Saclay, 61, avenue du président Wilson, 94235 Cachan Cedex, France.*

*Abstract.*

We report the use of high magnetomechanical coupling ferrites in magnetoelectric (ME) layered composites. Bilayer samples combining $(Ni_{0.973} Co_{0.027})_{1-x}Zn_xFe_2O_4$ ferrites (x=0-0.5) synthetized by non conventional reactive Spark Plasma Sintering (SPS) and commercial lead zirconate titanate (PZT) were characterized in term of ME voltage coefficients measured at sub-resonant frequency. Strong ME effects are obtained and we show that an annealing at 1000°C and a quenching in air improve the piezomagnetic behavior of Zn-rich compositions. A theoretical model that predict the ME behavior was developed, focusing our work on the demagnetizing effects in the transversal mode as well as the longitudinal mode. The model shows that: (i) high ME coefficients are obtained when ferrites with high magnetomechanical coupling are used in bilayer ME composites, (ii) the ME behavior in transversal and longitudinal modes are quite similar, and differences in the shapes of the ME curves are mainly due the demagnetizing effects, (iii) in the transversal mode, the magnetic field penetration depends on the ferrite layer thickness and the ME coefficient is affected accordingly. The two later points are confirmed by measurements on ME samples and calculations. Performances of the ME composites made with high magnetomechanical coupling ferrites are compared to those obtained using Terfenol-D materials in the same conditions of size, shape, and volume ratio. It appears that a ferrite with an optimized composition has performances comparable to those obtained with Terfenol-D material. Nevertheless, the fabrication processes of ferrites are quite simpler. Finally a ferrite/PZT based ME composite was used as a current sensor.

## I. Introduction

Magnetoelectric[1] (ME) materials which exhibit a coupling between magnetic properties and dielectric properties have a great interest due to their potential use in smart electronic applications. The ME effect consists in an electric polarization change when a magnetic field is applied (direct effect) or, conversely, a magnetization change when an electric field is applied (converse effect). Intrinsic (single phase) ME coupling was first detected in $Cr_2O_3$ at low temperature, but the effect is very weak and it cannot be used in electronic devices. More recently, both ferroelectric and ferromagnetic properties were discovered in bismuth ferrite[2] at room temperature, however the observed ME magnitudes are still too weak for application purposes. Recently, intrinsic ME effect suitable for some application has been reported in materials[3] derived from barium zirconium titanate with a small substitution of Fe. A possible alternative to the research of new intrinsic ME materials with enhanced effect is represented by the use of magnetostrictive/piezoelectric composites. In this case the ME effect arises through the mechanical coupling between the two phases[4]. When a constant magnetic bias field is applied to the composite, the magnetostrictive material shows a piezomagnetic behavior and the application of a small alternative field gives rise to a corresponding alternative strain. Due to mechanical coupling, an alternative electrical field appears within the piezoelectric phase. ME bulk composites have been made first by sintering together $BaTiO_3$ or PZT with spinel ferrites. However, the low resistivity of the ferrite that short-cut the piezoelectric phase acts towards a lowering of the ME effect[5]. This difficulty is overcome when using laminated structures[6].

So, in the 2000s, interest has been focused on layered composite structures based on PZT associated with different ferrimagnetic materials: nickel ferrites, cobalt ferrites, or lanthanum manganites[7]. Although nickel ferrites exhibit much smaller magnetostriction than cobalt ferrites, they show a better ME effect which is further strengthened by zinc substitution[7]. Furthermore, ME coupling seems to be dependent on geometric parameters such as layer thickness, number of layers, and volume ratio of the two phases. Strongest ME effects were obtained for co-sintered thin layers of PZT/nickel-zinc ferrite in the range of 10µm using the tape-casting[8] route. Giant magnetostrictive material such as Terfenol-D (Terbium Dysprosium iron alloy), were also used in laminated configurations associated with PZT or PMN-PT at resonant or sub-resonant frequencies[4,9,10]. Rare earth-iron alloys have been developed to reach large magnetostrictive strains (>1000ppm), and one obtains piezomagnetic coefficients as high as $d \sim 10^{-8} m/A$ at optimum bias field. Although giant ME coupling was obtained, the use of Terfenol-D is not suitable for the following reasons: (i) this material is extremely brittle and it cannot be machined easily; (ii) it cannot be co-sintered with piezoelectric materials; (iii) its high electric conductivity limits the use at low frequency because of the eddy currents; (iv) terbium and dysprosium is rare and expensive.

A piezoelectric/piezomagnetic bilayer can be modeled as two mechanically coupled transducers. In the 1950s and 1960s, some investigations on cobalt-substituted nickel ferrite transducers were conducted and some materials suitable for underwater applications were obtained[11,12,13]. Although the best composition of ferrite quenched in air achieved a coupling factor as high as 0.37,[12] in the 1960s research has turn towards PZT materials which have better electromechanical coupling. Nevertheless, in the field of ME composite, replacing conventional nickel ferrite with low magnetomechanical coupling factor[11] (k = 0.18) by cobalt-substituted nickel ferrite can lead to an improvement of the ME effect. Piezomagnetism is an intrinsic property which depends on the internal DC Bias and AC exciting magnetic fields. The demagnetizing field[14], depending on the overall geometry of the system, can greatly influence the ME response of a layered ME composite in the longitudinal as well as the transversal modes. Actually, a layered structure is strongly anisotropic and the demagnetizing field will affect rather differently the response of a system of given diameter and thickness under radial or axial excitations.

In the field of ME laminated composites, some studies have already be made on the influence of the demagnetizing effect (see Ref. 15, 16, and 17), but in these cases only the ME transversal mode was considered. However, here, we developed a simple model to take into account the influence of the demagnetizing field in the ME transversal as well as longitudinal modes for a layered structure. Especially, we have shown how the ME response in the longitudinal mode can be easily deduced from the measured response in the transversal mode (and vice-versa).

This paper is organized as follow. In section II, we present a model derived from the piezoelectric and piezomagnetic equations in a two dimensional quasi-static approach where demagnetizing corrections are accounted for each ME coupling mode (longitudinal and transversal), using the demagnetizing factor *N*. Useful tables of *N* values for discs or cylinders magnetized along radius or axis are available in the literature[18,19]. For a given sample, the links between the axial and radial ME responses are studied theoretically. For the transversal ME coupling, the effect of a decrease in the magnetic layer thickness was also theoretically studied in term of demagnetizing effect and field penetration.

In section III, we present the fabrication method of The ME samples. $(Ni_{0.973} Co_{0.027})_{1-x}Zn_xFe_2O_4$ compositions, with x = 0, 0.125, 0.25, 0.5, have been synthesized by reactive Spark Plasma Sintering (SPS) method[20]. To the best of our knowledge, concerning layered ME composites, no publication in the literature concerns ferrites made by reactive SPS methods. However, this method permits high densification and an increase of the magnetostrictive properties can be expected. Zinc substitution has been used to help the sintering stage with SPS to be accomplished because of the high diffusion capability of the $Zn^{2+}$ ions. Moreover, as showed by some authors[7], a certain amount of zinc permits to optimize the piezomagnetic behavior of ME composites. After SPS, all samples were annealed in natural atmosphere at 1000°C for re-oxidation, and some of them were quenched in air to estimate the effect of a heat treatment. All ferrite discs were pasted on commercially available PZT discs to form bilayers or trilayers and transversal ME coefficient measurements were performed at sub-resonant frequency (80Hz) .

In section IV, the ME responses of the ME samples made with ferrites with different Zn contents and heat treatments are presented. A possible explanation concerning the quenching effects is given. In this purpose, we have developed an original set-up to measure the intrinsic ME coefficients of ME samples. The aim of the measurement is to avoid the demagnetizing effect in the magnetic material when an AC magnetic field is applied. In the other hand, the link between the transversal and longitudinal ME modes is studied experimentally and theoretically in terms of demagnetizing effects and mechanical coupling. In the transversal mode, the demagnetizing effects in ME responses are also studied for samples including ferrite with various layer thicknesses and configurations. The mechanical coupling between ferrite and PZT layers is an important parameter that influence the ME response and it depends on the configurations of the layers. Studies were conducted on bilayers and trilayers to show what is the configuration with the best mechanical coupling. For a non-symmetrical sample, the possible effect of the flexural strain is bring out.

The performances of these ferrite/PZT bilayers have been assessed against other ME composites with high performances such as Terfenol-D/PZT bilayer. A Terfenol-D disc was purchased and pasted on a PZT disc. For relevant comparison, this ME sample exhibit the same geometry, dimension, and volume ratio, as our ferrite/PZT ME samples. The ME responses were explained estimating the piezomagnetic and elastic properties of both materials.

Finally, in section V, we report the fabrication of a passive current sensor prototype based on a transversal ME coupling. Layered ME samples made with NiCoZn-ferrites coupled with PZT exhibit a magnetic field sensitivity suitable to build non-resonant magnetic field or current sensors. It was characterized using square and triangle waveforms and it shows capabilities for high level AC current measurements ($I_{AC} \geq 100mA$) in the kHz range with good accuracy.

**II. Theoretical analysis.**

In order to explain the behavior of a layered ME composite, we have first developed a quasi-static 2-dimensional model, with particular attention towards the demagnetizing effects in longitudinal as well as transversal coupling modes. The interface of the bilayer is in the (1,2) plane and we assumed a perfect mechanical coupling between the two phases. The external mechanical influences on the ME composite are considered negligible. The induced electric field is always in the (3) direction. Transversal ME coupling refers to the case whence DC bias and AC magnetic fields are both applied in the direction (1). Longitudinal coupling refers to bias and AC magnetic field applied in the direction (3) (see Fig. 1).

*1. Transversal coupling.*

The following constitutive equations are used for the direct and converse piezoelectric effects (Einstein summation convention):

$$S_\alpha^e = s_{\alpha\beta}^E T_\beta^e + d_{i\alpha}^e E_i \qquad (1)$$

$$D_i = \varepsilon_{ij}^T E_j + d_{i\alpha}^e T_\alpha \qquad (2)$$

where $S_\alpha^e$ and $T_\beta^e$ are strain and stress tensor components, $D_i$ and $E_j$ are electric induction and electric field tensor components, and $s_{\alpha\beta}^E$, $d_{i\alpha}^e$, $\varepsilon_{ij}^T$ are zero field compliance, piezoelectric coefficient, and zero stress permittivity tensor components respectively.

In the same way, the magnetostrictive ferrite can be modeled as a piezomagnetic material, governed by the constitutive equations:

$$S_\alpha^m = s_{\alpha\beta}^H T_\beta^m + d_{i\alpha}^m H_i \qquad (3)$$

$$B_i = \mu_{ij}^T H_j + d_{i\alpha}^m T_\alpha \qquad (4)$$

where $S_\alpha^m$ and $T_\beta^m$ are strain and stress tensor components, $H_i$ and $B_i$ are internal magnetic field and internal magnetic induction (within the ferrite) tensor components, and $s_{\alpha\beta}^H$, $d_{i\alpha}^m$, $\mu_{ij}^T$ are zero field compliance, piezomagnetic coefficient (which depends on the magnetic bias), and zero stress permeability tensor components respectively.

In the bilayer configuration, the interface coincides with the 1-2 plane, and only extensional strain in those directions are considered. So only the $T_1^e, T_1^m, T_2^e, T_2^m, S_1^e, S_1^m, S_2^e, S_2^m$ tensor components are to be considered. The electric field is produced in the direction (3) only, so $E_i$ and $D_i$ reduced to $E_3$ and $D_3$. Lastly, assuming open electrical circuit (no current flowing from one electrode to the other), $D_3 = 0$. So Eq. (1) and (2) reduce to:

$$S_1^e = s_{11}^E T_1^e + s_{12}^E T_2^e + d_{31}^e E_3 \tag{5}$$

$$S_2^e = s_{12}^E T_1^e + s_{22}^E T_2^e + d_{31}^e E_3 \tag{6}$$

$$D_3 = d_{31}^e T_1^e + d_{31}^e T_2^e + \varepsilon_{33}^T E_3 = 0 \tag{7}$$

In the same way, Eq. (3) reduces to:

$$S_1^m = s_{11}^H T_1^m + s_{12}^H T_2^m + d_{11}^m H_1 \tag{8}$$

$$S_2^m = s_{12}^H T_1^m + s_{22}^H T_2^m + d_{12}^m H_1 \tag{9}$$

if the AC and DC magnetic field are applied along the direction (1).

At equilibrium, the force into the ferrite equates the one within the PZT (if we assume no forces absorbed by the glue layer), so from the stress point of view:

$$T_1^m = -\gamma T_1^e \tag{10}$$

$$T_2^m = -\gamma T_2^e \tag{11}$$

where $\gamma = {v_e}/{v_m}$, $v_e$ is the volume fraction of PZT and $v_m$ is the volume fraction of ferrite. Assuming ideal coupling at the interface, ferrite and PZT are equally strained, so $S_1^m$ equates with $S_1^e$ in one part, and $S_2^m$ equates with $S_2^e$ in the other part.

So, equating Eq. (5) with (8) on the one hand and Eq. (6) with (9) on the other, and writing the two expressions as a function of the stress within the PZT (using Eq. (10) and (11)), one obtains the formula:

$$[s_{11}^E + s_{12}^E + \gamma(s_{11}^H + s_{12}^H)](T_1^e + T_2^e) = (d_{12}^m + d_{11}^m).H_1 - 2d_{31}^e E_3 \tag{12}$$

Eq. (7) gives the electric field generated by the stress within the PZT:

$$E_3 = -\frac{d_{31}^e}{\varepsilon_{33}^T}(T_1^e + T_2^e) \tag{13}$$

Lastly, combining Eq. (12) and (13), the electric field $E_3$ is related to the AC internal magnetic field $H_1^{in}$:

$$E_3 = -\frac{d_{31}^e(d_{12}^m + d_{11}^m)}{\varepsilon_{33}^T[s_{11}^E + s_{12}^E + \gamma(s_{11}^H + s_{12}^H)] - 2(d_{31}^e)^2}.H_1 \tag{14}$$

This formulation for the transversal coupling is similar to the one given by some authors[21]. But in Eq. (14) the induced electric field is related to the internal magnetic field.

*2. Longitudinal coupling.*

In this configuration Eq. (5), (6), and (7), governing the piezoelectric material behavior, remain unchanged. In Eq. (8) and (9), the direction of the internal magnetic field must be changed from direction (1) to direction (3) and consequently, the piezomagnetic coefficient is now:

$$S_1^m = s_{11}^H T_1^m + s_{12}^H T_2^m + d_{31}^m H_3 \quad (15)$$

$$S_2^m = s_{12}^H T_1^m + s_{22}^H T_2^m + d_{31}^m H_3 \quad (16)$$

Lastly, the formula coupling the electric field $E_3$ and the AC internal magnetic field $H_3$ can be easily deduced from Eq. (14):

$$E_3 = -\frac{d_{31}^e(2d_{31}^m)}{\varepsilon_{33}^T [s_{11}^E + s_{12}^E + \gamma(s_{11}^H + s_{12}^H)] - 2(d_{31}^e)^2} \cdot H_3 \quad (17)$$

It must be noted that a flexural strain exists within an asymmetric bilayer, and it leads to a lowering of the extensional strain level. The model proposed here do not take into account the flexural behavior, and consequently the calculation may overestimate the ME response of a bilayer. Nevertheless, this model gives an overall trend of the mechanic, piezomagnetic, and piezoelectric parameter influences.

*3. Demagnetizing effect.*

Experimentally, the magnetoelectric effect is measured by applying to the ME sample a small AC magnetic field (produced by Helmholtz coils) superimposed to a DC field (produced by an electromagnet). The level of the applied AC field can be measured by means of a Hall effect probe or a search coil. Consequently, the ME coefficient is normalized using the applied magnetic field $H^a$. In order to compare theory and experiments, theoretical formulas (Eq. (14) and (17)) must be expressed in term of the external field $H^a$. The demagnetizing field $\vec{H}^d$ created by the sample is opposed to the external applied field $\vec{H}^a$ produced by the Helmholtz coils. Consequently the internal field $\vec{H}$ is lower than the external one. Within the magnetic material, the internal magnetic field[15] (in average over the volume) is:

$$\vec{H} = \vec{H}^a + \vec{H}^d = \vec{H}^a - N \cdot \delta \vec{M} \quad (18)$$

where $N$ is the magnetometric demagnetizing factor ($0 \leq N \leq 1$), and $\delta \vec{M}$ is the alternative magnetization variation produced by the small AC magnetic field. The alternative magnetization is related to the internal AC field:

$$\delta \vec{M} = \chi \vec{H} \quad (19)$$

Where $\chi$ is the reversible magnetic susceptibility (under constant stress). The internal AC field is expressed as:

$$\vec{H} = \frac{1}{1+N.\chi} \vec{H}^a \tag{20}$$

$\chi$ can be considered constant for $\delta M \ll M_s$ and practically independent of $H_{DC}$ if $H_{DC} < M_s$.

In the same way, the bias DC field produced by the electromagnet is reduced by the demagnetizing effect in the ferrite. The internal DC field is given by:

$$\vec{H}_{DC} = \frac{1}{1+N.\chi_{DC}} \vec{H}^a_{DC} \tag{21}$$

Where $\chi_{DC} = M/H_{DC}$ is the static magnetic susceptibility (under constant stress).

Consequently, the ME coefficient with respect to the external applied field, for the transversal coupling mode, $\alpha_{31}$, and for the longitudinal one, $\alpha_{33}$, are given by:

$$\alpha_{31} = \frac{E_3}{H_1^a} = -\frac{d_{31}^e(d_{12}^m+d_{11}^m)}{\varepsilon_{33}^T[s_{11}^E+s_{12}^E+\gamma(s_{11}^H+s_{12}^H)]-2(d_{31}^e)^2} \cdot \frac{1}{1+N_r\chi} \tag{22}$$

$$\alpha_{33} = \frac{E_3}{H_3^a} = -\frac{d_{31}^e(2d_{31}^m)}{\varepsilon_{33}^T[s_{11}^E+s_{12}^E+\gamma(s_{11}^H+s_{12}^H)]-2(d_{31}^e)^2} \cdot \frac{1}{1+N_z\chi} \tag{23}$$

Where $N_r$, and $N_z$ are the radial and axial magnetometric demagnetizing factors respectively.

In the case of cylinders, the demagnetizing factors $N_r$ and $N_z$ depend on the ratio of thickness to diameter $t/d$ of the magnetic sample and little on the susceptibility $\chi$ of the material. In Fig. 2, using data published by D. X. Chen et al.,[18,19] the radial and axial demagnetizing factors were plotted as function of $t/d$ for different values of susceptibility. Good accuracy are obtained for demagnetizing factor $N_r(t/d)$ and $N_z(t/d)$ expressed as function of thickness to diameter ratio only. For a given piezoelectric material, mechanical, dielectric, and piezoelectric properties can be regarded as constants. So, the shape of the ME curves depend only on the intrinsic piezomagnetic coefficients and the static and reversible susceptibility $\chi(H_{DC})$ which are controlled by the internal bias field $H_{DC}$. Note that the intrinsic piezomagnetic coefficients are related to strain derivatives with respect to the internal field $H$. According to Eq. (20), and (21) demagnetizing effects also influence the ME bilayers behavior.

Eq. (22) and (23) are quite similar. They differ only in the piezomagnetic coefficients and the demagnetizing factors. We can expect similar magnetoelectric behaviors when measurement are made in the transversal and longitudinal modes. Because Joules magnetostriction is a constant volume deformation (in general, volume magnetostriction can be neglected), for polycrystalline ferrite with random lattice orientations, the piezomagnetic coefficients are simply linked and it can be assumed that $d_{11}^m \approx -2\, d_{12}^m \approx -2\, d_{31}^m$ for the same internal bias field.

Theoretically, according to Eq. (22) and (23), high magnetoelectric effects are obtained when the piezomagnetic material has the following properties: (i) low reversible magnetic susceptibility $\chi$ which permits high internal AC field, (ii) high intrinsic piezomagnetic coefficient $d$, (iii) low compliance coefficients $s_{11}^H$ and $s_{12}^H$. It is equivalent to maximize the following ratio: $\frac{d}{s^H \chi}$. This ratio is quite similar to the magnetomechanical coupling factor[22] $k = \frac{d}{\sqrt{s^H \mu}}$ (where $\mu = \mu_0(1+\chi)$ is the absolute permeability) which indicates the capabilities of a magnetostrictive material to convert magnetic energy to mechanical one (and vice-versa). In conclusion, good magnetic materials for magnetoelectric applications are to be chosen among the class of high magnetomechanical coupling piezomagnetic ferrite commonly used in ultrasonic transducer applications. In the 50'S and 60s, many investigations[11,12,13] have been made in the field of underwater acoustic. It turns out that cobalt substituted nickel ferrites have high magnetomechanical coupling factor, within the order of magnitude of 0.3 (at the optimum bias). These compositions will be the starting point for our study of ME bilayer.

### III. Experiment.

*1. Samples fabrication.*

Nanosized (<50nm) powders purchased from Sigma-Aldricht ($Fe_2O_3$, $NiO$, $Co_3O_4$, $ZnO$) were used as precursor oxides. The compositions ($Ni_{0.973}$ $Co_{0.027})_{1-x}Zn_xFe_2O_4$, where x = 0, 0.125, 0.25, 0.5, were obtained by mixing the different powders in appropriate ratios. All powders were mixed by ball milling (30 min at 200 rpm), and then grinded at 600 rpm for 1 hour. SPS was used for the spinel structure formation and sintering stages. Mixtures were loaded in a cylindrical graphite die (10mm inner diameter) between two pistons applying uniaxial stress. Carbon paper was used as separator between powder, carbon die and pistons. The reaction and then the sintering were carried out in neutral atmosphere (argon) by heating the powder with a pulsed DC current under application of uniaxial stress. The reaction stage was performed at 600°C under a pressure of 100MPa for 5 minutes. The densification is obtained at 850°C for 3 minutes under the same pressure. After sintering, the samples are obtained in the form of 10mm diameter and 2mm thick discs. For each composition, several samples were produced. All these samples were annealed in air at 1000°C for 1 hour. For each composition, we produced two samples: one quenched in air (except x=0) and the other slowly cooled (approximately 3.5°C/min). Part of ferrite discs (2mm thickness) were directly pasted with silver epoxy (Epotek E4110) on PZT discs (PZ27 (Ferroperm)) with the same diameter. One other part of ferrite discs were cut to reduce the thickness (1mm) and pasted on PZT discs (PZ27). Finally, two ferrite discs were drilled to form rings, and then, they were pasted on PZT rings (Pic255 (Physik Intrumente)). All the piezoelectric discs or rings are polarized in the thickness direction (3). The ME sample characteristics are resumed in Table 1. It must be noted that the SPS sintering process in argon atmosphere produces ferrites with some deficiency in oxygen, and some $Fe^{3+}$ ions reduce into $Fe^{2+}$. The annealing at 1000°C in natural atmosphere do not re-oxidize the ferrite completely because of the low porosity. As a consequence, the conductivity of the material increases, but this is not affecting layered ME composites performances.

*2. ME measurement procedure.*

The ME coefficient is measured as a function of external applied DC magnetic field $H_{DC}^a$ produced by an electromagnet[23]. For each $H_{DC}^a$ working point, a small external applied AC magnetic field $H^a$ (1mT, 80Hz) produced by Helmholtz coils is superimposed. Low frequency is used to avoid any resonance effect, so ME bilayers are excited in quasi-static mode. The ME voltage is measured by means of a lock-in amplifier (EG&G Princeton Applied Research Model 5210) with high input impedance (100MΩ). All the measurement were performed at room temperature. When the external magnetic fields (AC and DC bias) are perpendicular to the electrical polarization direction, the experimental transversal ME coefficient is :

$$\alpha_{31} = \frac{E_3}{H_1^a} \quad (24)$$

where $E_3$ is the electric field within the PZT layer. $H_1^a$ is the external AC field applied in the direction (1). When the magnetic fields (AC and DC bias) are parallel to the electric polarization direction, the experimental longitudinal ME coefficient is :

$$\alpha_{33} = \frac{E_3}{H_3^a} \quad (25)$$

where $H_3^a$ is the external AC field applied in the direction (3).

## IV. Results and discussion.

### 1. Influences of the quenching and Zn contents in the ferrites.

In order to measure the effects of heat treatment and Zn content on the properties of ferrites, the transversal ME coefficients were recorded for the different ferrite compositions (x= 0, 0.125, 0.25, and 0.5), for slow-cooled and quenched samples (see Fig. 3). The maximum ME effect occurs at magnetic DC fields between 30 kA/m and 55kA/m for all compositions. The maximum ME effect is obtained for x=0.125, and decreases with Zn content. For x=0.125, the ME effect is larger for the slow-cooled samples. By opposition, for x=0.25 and 0.5, the quenching enhances the ME coefficients. The increase in the non-magnetic $Zn^{2+}$ content from x=0.125 to x=0.5 produces two effects which work together: (i) the saturation magnetostriction decreases because of the dilution of the magnetostrictive cations ($Ni^{2+}$, $Co^{2+}$, $Fe^{2+}$), and so, the piezomagnetic coefficient decreases too; (ii) the permeability increase and as a consequence the internal field decreases according to Eq. (20). A possible explanation of the effect of heat treatment can be given as follow. $CoFe_2O_4$ is usually considered as an inverse spinel ferrite, but the cations distribution depends on the heat treatment[24]. At high temperature, $Co^{2+}$ and $Zn^{2+}$ ions which have the same radius (82 pm) may be distributed in both sites (tetrahedral and octahedral) because magnetic interaction have no effect.  By quenching, a part of $Co^{2+}$ cations can be frozen into their high temperature positions which correspond to a more randomized distribution[25]. By opposition, slow-cooling gives time for cations to move into their low temperature equilibrium positions. In contrast, for $NiFe_2O_4$ ferrite, the cations distribution (inverse spinel) seems to be independent of the heat treatment or cooling method of the fabrication process[24]. In case of Co substituted NiZn-ferrites, a reasonable assumption is that the quenching affects the distribution of the $Co^{2+}$cations in the tetrahedral and octahedral sites. This distribution is also influenced by the $Zn^{2+}$ rate: $Zn^{2+}$ cations which occupy the tetrahedral sites can affect the migration of $Co^{2+}$ cations during a thermal treatement. Thus, quenching and $Zn^{2+}$ rate affects the local electronic symmetry, and in turn, spin-orbit coupling dependent properties, namely the magneto-crystalline anisotropy and the magnetostriction.

## *2. Link between transversal and longitudinal modes.*

To investigate the effects of the coupling modes, transversal and longitudinal ME coefficients were measured (see Fig. 4) for the sample # 0.5SC2/1. As expected, for longitudinal coupling, for which the AC and DC demagnetizing fields are higher, the peak of the ME curve is lower and it occurs at higher external bias field. Eq. (22) and (23) show that transversal and longitudinal ME coefficients are strongly linked and that they can be deduced from each other if demagnetizing field corrections are done, knowing the static and reversible magnetic susceptibility of the ferrite. For this purpose, a spherical sample of ferrite was machined (diameter: 2mm) and characterized using a Vibrating Sample Magnetometer (VSM) (LakeShore 7404). The virgin magnetization curve versus internal field (after demagnetizing field correction) is given Fig. 5. From this measurement, the differential susceptibility $dM/dH$ was deduced and plotted on the same figure. From transversal coupling to longitudinal coupling, due to the demagnetizing effects, the amplitude of the ME response must be corrected by the ratio $(1 + N_r\chi)/(1 + N_z\chi)$. Obviously, at low magnetic bias, it is known that the reversible susceptibility is different from the differential susceptibility. But in this case, where both reversible and differential susceptibility are high, the ratio is independent of the susceptibilities and can be approximated by $N_r/N_z$. The value of the susceptibility affects the result only if $N.\chi$ is within the order of 1 which practically means $\chi \lesssim 20$ in the present case (or for $H$ above 50kA/m). In this region (approach to saturation) the differential and reversible susceptibilities are equal. According to Eq. (20), (21), (22), and (23), the solid line curve in Fig. 4 is deduced from the experimental transversal coupling curve (dashed line), making demagnetizing field corrections. It is noted that for isotropic polycrystalline ferrites, $d_{31}^m(H_{DC}) \sim d_{11}^m(H_{DC}) + d_{21}^m(H_{DC})$ in theory. Here, to fit the curves, we have taken piezomagnetic coefficients with a ratio of $d_{31}^m/(d_{11}^m + d_{21}^m) \sim 0.63$. This deviation can be explained by an anisotropic piezomagnetic behavior of the ferrite. This point will be discussed in section IV.5. of the paper. Concerning the transversal coupling, the peak of the ME coefficient is obtained for a $4 \times 10^4 A/m$ external applied bias field. At this working point, the reversible susceptibility is around 23 and according Eq. (20), the internal AC field is four times lower than the external one. Hence, due to the demagnetizing field, the ME coefficient is correspondingly diminished.

## *3. Ferrite layer thickness effects on the ME response.*

As seen before, radial and axial demagnetizing effects, which are quite different for a given sample, have a considerable influence on the ME response. The demagnetizing effect can be tuned in the same radial mode by changing the thickness of the magnetic material in the ME samples: according to Fig. 2, the radial demagnetizing factor *$N_r$* decreases when the thickness is decreased for a given diameter. To illustrate this point, a ME bilayer sample was made with a thin ferrite disc of 1mm in thickness pasted on a 0.5mm PZ27 disc (sample #0.125SC1/0.5). Its behavior was compared to the reference ME sample with 2mm thick ferrite layer of the same composition pasted on a 1mm PZ27 disc (sample # 0.125SC2/1). Note that for better comparison, the two ME samples have the same PZT/ferrite volume ratio ($\gamma=1/2$). The two measured ME coefficients are given in Fig. 6. Both curves have approximately the same amplitude whereas the main effect of the demagnetizing field is to shift the peaks of the curves. The experimental ME curve of the reference sample (# 0.125SC2/1) was corrected in demagnetizing effect to obtain the behavior of the 1mm thick ferrite ME sample. The method used here is similar to the one used before for the longitudinal ME coefficient calculation: a spherical shape of ferrite was characterized using a VSM. From the virgin magnetization curve, the susceptibilities were deduced, and then demagnetization corrections were done. This theoretically corrected curve (thin dashed line in Fig. 6) exhibits correct downshift in the peak position but the ME amplitude is slightly overestimated with respect to the measured one. Next, a trilayer ferrite/PZT/ferrite of the same composition with 1mm thick for each ferrite layer pasted on a 1mm thick PZ27 layer was made (sample # 0.125SC1/1/1). Its ME coefficient (plotted in Fig. 6, thick dotted line) is quite higher than that of the reference bilayer sample. This behavior cannot be explained by a demagnetizing effect. In fact, the corrected curve in demagnetizing effect (thin dotted line) from that of the reference sample have quite lower amplitude compared to the experimental one. We assume that this high ME effect is due to a higher mechanical coupling between PZT and ferrite layers because the PZT layer is stressed on its two faces. Furthermore, due to the symmetric geometry of the trilayer, no flexural strain occurs: a higher stress is transmitted from the two ferrite layers to the piezoelectric one. In this case, the mechanical behavior of the ME composite is closer to the simplified model calculated in section II (see section IV. 5. for precisions) Note that for the trilayer sample, the horizontal shift in the ME curve is half less than the bilayer one. Whereas in the two cases the ferrite layers have the same thickness (1mm), in the trilayer case there is a magnetic coupling between the two ferrite layer, increasing the demagnetizing factor. We estimate that this factor is increased from 0.0881 for a single ferrite layer (1mm thick) to 0.115 for two ferrite layers separated by a 1mm PZT layer. To confirm the overall trend of a ME trilayer, one another sample (# 0.125SC0.5/0.5/0.5) was made: two ferrite layer (0.5 mm thick each) were pasted on both faces of a PZ27 layer (0.5mm thick). Its measured ME coefficient is plotted (dashed line) in Fig. 7. Due to a lower demagnetizing field and a good mechanical behavior, the peak ME coefficient is increased up to 0.55 V/A.

*4. Effect of an inhomogeneous strain on the ME response.*

The trilayer sample (# 0.125SC1/1/1) have quite higher ME coefficient compared to the bilayer reference sample (# 0.125SC2/1), and the whole difference cannot be explain if we consider only the demagnetizing effect. Firstly, we may assumed that a flexural strain of the bilayer produces a decrease of the stress in the piezoelectric layer compared of the case of a trilayer with no flexural strain. To answer this question, a trilayer sample was made: two PZ27 layers (0.5mm thick each) were pasted on both faces of a ferrite layer (2mm thick). This sample (# 0.125SC0.5/2/0.5) have the same demagnetizing factor and the same volume ratio as the reference sample. But due to its symmetric geometry, no flexural behavior can occur. The

ME coefficient was measured and plotted in Fig. 7 (dotted-dashed line). This trilayer sample and the bilayer (reference) sample have comparable ME behavior with the same peak value of 0.2V/A. We can conclude that the flexural strain behavior of the reference bilayer sample have negligible influence on the peak of the ME coefficient since the whole thickness (3mm) is only 3.3 times lower than the diameter of the sample (10mm). Lastly, only the in-plane strain have to be considered to predict the ME response. The exact solution of the strain field (strongly inhomogeneous) in the whole ME sample can be calculated only by numerical methods (Finite Element Method, for example). Nevertheless, an experimental estimation of the strain level in each layers of the ME sample can be made. For this purpose, two strain gauges (EA-06-062TT-120, Micro-Measurements) were pasted on each free faces of the PZT and ferrites layers of the reference bilayer sample (# 0.125SC2/1). The strain gauges were chosen because of their small sensitive areas (1.8×1.8 mm$^2$) that permits local measurements near the edge of the faces where the strains are expected to be highly different. The longitudinal strains versus DC magnetic field are given in Fig. 8. As expected, the strain at the surface of ferrite layer is quite higher than the one measured for the PZT layer. At saturation, we obtain $S_1^m \approx -34 ppm$ for the ferrite, and $S_1^e \approx -7 ppm$ for the PZT. Although these measurements were made on surfaces, average values of the strains in each layers can be estimated when a linear profile for the strain along the direction (3) is assumed. Using geometric considerations, we obtain $\langle S_1^m \rangle \approx -28.5 ppm$ in the ferrite and $\langle S_1^e \rangle \approx -11.5 ppm$ in the PZT. The ratio of strains, $\eta = \langle S_1^e \rangle / \langle S_1^m \rangle \approx 0.4$ at saturation, is quite low. In fact, the stress is applied on one face of the PTZ layer, the other face staying mechanically free. So, the propagation of a longitudinal strain from the ferrite/PZT interface to the free face of the PZT layer is mainly due to a shear stress effect. Note that the value of the ferrite strain at saturation ($S_1^m \approx -34 ppm$) of the ME sample is higher than the magnetostriction saturation ($\lambda_{11}^s = -26 ppm$) measured for a mechanically free material (see Fig. 13). This difference can be explained as following: (i) the strain gauge was pasted near the edge of the ferrite disc of the ME sample while the magnetostriction measurement was done with a strain gauge centered on the surface of the disc where the strain is theoretically lower; (ii) the PZT layer acts like a mechanical load (or pre-load) on the ferrite layer, and as a consequence, the magnetostriction at saturation is increased in comparison to a mechanically free ferrite sample. The average strains in ferrite layer and PZT layer are quiet different: we have estimated $\langle S_1^e \rangle = 0.4 \langle S_1^m \rangle$ and this effect highly deceases the ME response of a bilayer sample. So when the strain ratio is introduced for the longitudinal mechanical coupling, $\eta = \langle S_1^e \rangle / \langle S_1^m \rangle$, as well as for the transversal mechanical coupling, $\eta = \langle S_2^e \rangle / \langle S_2^m \rangle$, with the same value $\eta$, Eq. (22) is modified as (transversal ME coupling):

$$\alpha_{31} = \frac{E_3}{H_1^a} = -\frac{\eta d_{31}^e (d_{12}^m + d_{11}^m)}{\varepsilon_{33}^T [s_{11}^E + s_{12}^E + \eta \gamma (s_{11}^H + s_{12}^H)] - 2(d_{31}^e)^2} \cdot \frac{1}{1 + N_r \chi} \quad (26)$$

Obviously, an equivalent modification can be applied to Eq. (23) for the longitudinal ME coupling.

## 5. Effect of the quenching on the ME response.

For Zn-rich samples (x=0.25 and x=0.5), the quench of ferrites permits to enhance the magnetoelectric coupling. According Eq. (22), this enhancement can have two main causes: the improvement of the intrinsic piezomagnetic coefficients or a weakening of the magnetic susceptibility. To answer this question, the following experiments were made. For the composition with x=0.25 (quenched and slow-cooled), the ferrite discs were delaminated from the PZ27 layers. Then, they were drilled to form ferrite rings with a centered 3 mm hole. A thin PZT ring (PIC255, 10mm outer diameter, 5mm internal diameter, 0.25mm thickness) was pasted with silver epoxy on each ferrite rings. Samples are referenced 0.25Q2/0.25 and 0.25SC2/0.25. A 8 turns coil was wound on each ME ring. AC tangential magnetic field $H_{AC}$ (in the (1,2) plane) is forced within the ferrite when the AC current, $I_{AC}$, is flowing into the coil: $H_{AC} = nI_{AC}/\pi/D$, where D=6.5mm, is the mean diameter of the ferrite ring, and n=8, is the number of turns (so the demagnetizing factor $N$ is 0 for $H_{AC}$). The top electrode of the PZT layers was separated into two parts (two strokes of file in the direction (1)) and then, the AC voltage produced by a half piezoelectric layer (parallel to the magnetic field) was measured by means of a lock-in amplifier. The ME samples were placed in a DC magnetic field applied in the direction (1) and produced by an electromagnet (see Fig. 9). The corresponding ME coefficients versus external applied field $H_{DC}$ are given in Fig. 10. The two curves, for quenched ferrite sample and slow-cooled ferrite sample, are quite similar: the same maximum amplitudes are obtained for close DC fields ($2 - 2.4 \times 10^4 A/m$). So the quench do not modify the intrinsic piezomagnetic behaviors of the ferrites. The differences obtained in ME curves for quenched and slowly cooled ferrite, when the ME coefficients are normalized using external applied AC field, are assumed to be the result of the demagnetizing effects. The initial permeability was measured for the two ferrites: $\chi = 100$ (quenched) and $\chi = 409$ (slow-cooled) and it confirms that a less penetrating magnetic field in the second case reduces the ME coefficient according Eq. (22).

*6. Co-substitued Ni-Zn ferrite properties compared to those of Terfenol-D.*

In a aim of comparison, Terfenol-D discs with the same size as the NiCoZn-ferrites (10 mm diameter and 2 mm thickness) were purchased from Etrema. One of them was pasted on a Pz27 disc (10mm diameter, 1mm thickness) with silver epoxy. The measured transversal ME coefficient is plotted in Fig. 11 and compared to that of our best bilayer sample # 0.125SC2/1. As expected, Terfenol-D produces better ME effects but it is only 1.8 times higher. This good performance of the ferrite/PZT composite compared to the Terfenol-D one should be explained estimating the piezomagnetic and elastic properties of both materials. Magnetostriction curves (parallel and perpendicular to the H field in the (1,2) plane of the sample) were measured using a strain gauge (H06A-AC1-125-700, Micro-Measurements) pasted on a Terfenol-D disc. $\lambda_{11}$ is the magnetostriction measured in the direction (1) when an external field is applied in the direction (1). $\lambda_{12}$ is the magnetostriction measured in the direction (1) when an external field is applied in the direction (2). The result is given Fig. 12. One observes saturation magnetostriction as high as $\lambda_{11}^s = 1000 ppm$ in the parallel mode and $\lambda_{12}^s = 700 ppm$ in the perpendicular one, and it is in the range of the data given by the manufacturer. At the optimal external bias field of $H_0^a = 10^5 A/m$ the piezomagnetic coefficients are $d_{11}^a = (d\lambda_{11}/dH_1^a)_{H_0^a} \sim 11.2 \times 10^{-9} m/A$ and $d_{12}^a = (d\lambda_{12}/dH_2^a)_{H_0^a} \sim -$

$7.3 \times 10^{-9} m/A$. It should be also noted that, in this particular case, the piezomagnetic coefficients $d_{11}^a$ and $d_{12}^a$ are related to strain derivatives with respect to the external applied field $H^a$, so the effects of the demagnetizing field is already contained in these measurements. In the same way, a strain gauge was pasted on a NiCoZn-ferrite (with x=0.125), and parallel and perpendicular in plane magnetostrictions were measured (see Fig. 13). The parallel magnetostriction saturation is $\lambda_{11}^s = -26 ppm$ which is approximately two times higher than the perpendicular one ($\lambda_{12}^s = +12 ppm$). One obtains the extrinsic piezomagnetic coefficients expressed as a derivative with respect to the external field: $d_{11}^a = (d\lambda_{11}/dH_1^a)_{H_0^a} \sim -0.67 \times 10^{-9} m/A$ and $d_{12}^a = (d\lambda_{12}/dH_2^a)_{H_0^a} \sim 0.32 \times 10^{-9} m/A$ at the optimal bias of $H_0^a = 3.75 \times 10^4 A/m$. The $s_{33}^H$ and $s_{31}^H$ compliances were deduced from ultrasonic velocity measurements along the thickness direction of a NiCoZn-ferrite disc by using a pulse-echo technique (shear and longitudinal waves) at 20MHz. The longitudinal ($V_l = 6200 m/s$) and transverse ($V_t = 3400 m/s$) waves velocities yield: $s_{33}^H = 6.47 \times 10^{-12} m^2/N$, and $s_{31}^H = -1.84 \times 10^{-12} m^2/N$. Assuming perfect isotropic polycrystalline ferrite, $s_{11}^H$ and $s_{12}^H$ reduce to: $s_{11}^H = s_{33}^H$ and $s_{12}^H = s_{13}^H$. The Terfenol-D disc purchased from Etrema is grain oriented in the thickness direction and exhibit an axial symmetry. As a consequence, in the transverse direction, the compliances are rather high: $s_{11}^H = 125 \times 10^{-12} m^2/N$ and $s_{12}^H = -37.5 \times 10^{-12} m^2/N$ (data from Etrema and Dong et al.[10]). So, we can say that the advantage due to the high piezomagnetic coefficient of Terfenol-D (17 times higher than the NiCoZn-ferrite one) is counter balanced by its high compliance (19 times higher than the NiCoZn-ferrite one). So according Eq. (22), one can roughly estimate that Terfenol-D and NiCoZn-ferrite have ME performance within the same order when associated with PZT.

The magnetostriction characterization is a helpful tool to estimate the magnetomechanical anisotropy of samples. For the ferrite, the parallel saturation magnetostriction, $\lambda_{11}^s = -26 ppm$, is two times higher (in amplitude) than $\lambda_{12}^s$. This leads to extrinsic piezomagnetic coefficients $d_{11}^a$ and $d_{12}^a$ in the same ratio, and consequently, the intrinsic piezomagnetic coefficients $d_{11}$ and $d_{12}$ have the same ratio. We can conclude to isotropic piezomagnetic behavior in the (1,2) plan. Measurement of saturation magnetostriction in the (1) direction, when a magnetic field is applied in the (3) direction, gives $\lambda_{13}^s = 7 ppm$, which is 4 times lower than $\lambda_{11}^s$. It can be assumed that the value of the intrinsic piezomagnetic coefficients $d_{13}$ is roughly $d_{13} \sim -0.25\, d_{11}$. As a consequence, the value of the ratio $d_{31}/(d_{11} + d_{12})$ is -0.5 which is quite different from -1, the theoretical value for a perfect isotropic ferrite. So, we can conclude to large anisotropic behavior for 3-1 piezomagnetic coupling. The physical reason of this phenomenon is unclear, but we assumed that a preferred direction is induced when the uniaxial stress is applied in the (3) direction during the reactive and sintering stages of the SPS synthesis.

Using data of piezomagnetic, piezoelectric, and elastic properties (summarized in Table 2), theoretical ME coefficients in transversal coupling at optimal bias were theoretically calculated. For sample # 0.125SC2/1, the theory given by Eq. (22) predicts $\alpha_{31} = 0.41\, V/A$ which is two times higher than the experimental one. Nevertheless, when a strain ratio of $\eta = \langle S_1^e \rangle / \langle S_1^m \rangle \approx 0.4$ (as measured before) is introduced, the theory predicts (Eq. (26)) $\alpha_{31} = 0.196\, V/A$ which is quite similar to $\alpha_{31} = 0.2\, V/A$ obtained experimentally. In contrast, for a trilayer of the same composition (sample # 0.125SC1/1/1) we obtain experimentally $\alpha_{31} = 0.36 (V/A)$ which is closer to the theoretical value (corrected in demagnetizing field) given by Eq. (22) $\alpha_{31} = 0.47 (V/A)$. Using Eq. (26), experiment and theory give the same result when a strain ratio of $\eta = \langle S_1^e \rangle / \langle S_1^m \rangle = 0.71$ is introduced. At this point, it is clearly seen that a piezoelectric layer stressed on its both faces by two ferrite layers produces average strain 77.5% higher compared to the case of a PZT layer stressed only on one face by a single ferrite layer. Concerning the Terfenol-D/P27 bilayer, theory (Eq. (22)) leads to $\alpha_{31} = 0.82 (V/A)$ which is little bit more than twice higher than the experimental result. Here again, this discrepancy is assumed to be due to a low average strain ratio $\eta$ when the piezoelectric layer is stressed by a single magnetostrictive layer. Using Eq. (26), theory and experiment give the same result if a strain ratio of $\eta = \langle S_1^e \rangle / \langle S_1^m \rangle = 0.1$ is introduced. This low strain ratio is assumed to be the result of the high stiffness of the Terfenol-D material.

From another point of view, the dynamic magnetostrictive properties of a material can be expressed in terms of energy conversion capabilities. The magnetomechanical coupling factor $k_{11} = d_{11}/\sqrt{s_{11}^H \mu}$ is a function of the intrinsic piezomagnetic coefficient $d_{11}$, related to the internal field, $d_{11} = (d\lambda_{11}/dH_1)_{H_0}$. This later can be expressed using the extrinsic piezomagnetic coefficient $d_{11} = d_{11}^a (1 + N_r . \chi)$ and the magnetomechanical coupling factor become $k_{11} = d_{11}^a (1 + N_r . \chi)/\sqrt{s_{11}^H \mu}$. Using data summarized in Table 2, calculations lead to $k_{11} = 0.35$ for the NiCoZn ferrite, and $k_{11} = 0.66$ for the Terfenol-D material which is roughly 2 times higher. It is interesting to see that the two ME coefficients of the two materials have the same ratio 2:1. This suggests that the magnetomechanical coupling factor is an appropriate criterion to select magnetostrictive materials to be used into ME multilayer composites.

**V. Sensor prototype.**

Lastly, we have made a prototype of a current sensor based on a ME composite. The trilayer ferrite/PZT/ferrite (sample # 0.125SC1/1/1) with $\alpha_{31} = 0.36 V/A$ was placed in a small AC-coil made with four turns of an isolated wire. The low bias field required to polarize the ferrite material is produced by two small permanent barium ferrite magnets. The distance between the magnets and the trilayer was set to obtain the highest ME effect. The AC current flowing into the AC-coil was produced by a power amplifier connected in series with a 4Ω power resistor. This current was sensing by a commercial active current probe (Tektronic A622) for comparison. The ME voltage was sensed directly by a passive voltage probe with 10MΩ input impedance connected to an oscilloscope (Lecroy Waverunner 44Xi). 10A peak square and triangle current waveforms at 1kHz were used to characterize the response of the ME sensor in the sub-resonant mode. Fig. 14 and Fig. 15 show that, at this frequency, the phase shift is negligible, and the sensor have a good linearity and low noise level. Here, the gain of the ME current sensor is approximately 87mV for 1A flowing into the wire. Obviously, this gain can be set to a given value by adding or removing some turns to the AC coil or by changing its size. We have already properly characterized current sensors with the same structure (but with lower sensibility). The results will be given in a paper[28] (not yet published).

## VI. Conclusion

Co-substituted Ni-Zn ferrite were synthesized using a non conventional solid state reaction and sintering route based on a SPS method. These ferrites showing high magnetomechanical coupling factors are suitable to make multilayer ME composites exhibiting strong ME effects as predicted by a theoretical model. ME performances obtained with an optimized composition of NiCoZn-ferrite/PZT bilayer are comparable to those obtained with the same structure combining Terfenol-D and PZT. It was shown that the low piezomagnetic coupling performance of the ferrite (in comparison to the Terfenol-D) is balanced by a low compliance that permits high stresses produced by low strains. Another important parameter that influence the ME behavior is the demagnetizing factor associated with the susceptibility of the magnetic material. By lowering the internal AC and bias magnetic fields, demagnetizing effects partially shape the ME coefficient curves. By varying the thickness of the magnetic layer or by changing the exciting modes (axial or radial), the peak of a ME coefficient curve can be tuned, and its amplitude is affected. In the axial mode, due to better penetration of the magnetic field, the theory predicts higher ME coupling for low thickness to diameter ratios and as a consequence, the ME curve peak is shift down. In conclusion, the ME behavior of a layered ME sample is the result of three intrinsic parameters of the magnetic material: the intrinsic piezomagnetic coupling coefficients, the compliance, and the magnetic susceptibility. Furthermore, it was shown that the demagnetizing factor and the mechanical coupling between the layers are the extrinsic parameters of the magnetic material that influence the ME behavior. Finally, we used Co-substituted Ni-Zn ferrite associated with PZT in a trilayer structure as a ME current sensor. The device was purely passive: the bias field was obtained with two barium ferrite magnets and the ME voltage was directly measured with a passive voltage probe. The characterization in a nonresonant mode have shown a high linearity and a negligible phase shift. The gain of the sensor is closed to 90mV/A for courant in the 0-10A range that meets performance suitable in the field of power electronic applications.

| Sample # | Ferrite Zn content | Layer configurations | Layer thickness (mm) | Ferrite thermal treatment |
|---|---|---|---|---|
| 0.5Q2/1 | x=0.5 | Ferrite/PZ27 | 2/1 | quenched |
| 0.5SC2/1 | x=0.5 | Ferrite/PZ27 | 2/1 | slow cooled |
| 0.25Q2/1 | x=0.25 | Ferrite/PZ27 | 2/1 | quenched |
| 0.25SC2/1 | x=0.25 | Ferrite/PZ27 | 2/1 | slow cooled |
| 0.125Q2/1 | x=0.125 | Ferrite/PZ27 | 2/1 | quenched |
| 0.125SC2/1 | x=0.125 | Ferrite/PZ27 | 2/1 | slow cooled |
| 0SC2/1 | x=0 | Ferrite/PZ27 | 2/1 | slow cooled |
| 0.25Q2/0.25 | x=0.25 | Ferrite/Pic255 | 2/0.25 | quenched |
| 0.25SC2/0.25 | x=0.25 | Ferrite/Pic255 | 2/0.25 | slow cooled |
| 0.125SC1/0.5 | x=0.125 | Ferrite/PZ27 | 1/0.5 | slow cooled |
| 0.125SC1/1/1 | x=0.125 | Ferrite/PZ27/Ferrite | 1/1/1 | slow cooled |
| 0.125SC0.5/0.5/0.5 | X=0.125 | Ferrite/PZ27/Ferrite | 0.5/0.5/0.5 | slow cooled |
| 0.125SC0.5/2/0.5 | X=0.125 | PZ27/Ferrite/PZ27 | 0.5/2/0.5 | slow cooled |

Table 1: Characteristics of the ME composite samples. The different ferrite compositions are $(Ni_{0.973} Co_{0.027})_{1-x} Zn_x Fe_2 O_4$.

|  | $d_{31}^e$ (pC/N) | $d_{11}^m$ (nm/A) | $d_{12}^m$ (nm/A) | $s_{11}^E$ or $s_{11}^H$ (m²/N) | $s_{12}^E$ or $s_{12}^H$ (m²/N) | $\mu^T$ or $\varepsilon_{33}^T$ (in relative) |
|---|---|---|---|---|---|---|
| PZ27 | -170 | | | $17 \times 10^{-12}$ | $-6.6 \times 10^{-12}$ | 1800 |
| Pic255 | -180 | | | $15 \times 10^{-12}$ | - | 2400 |
| NiCoZn-ferrite | | -0.67 | 0.32 | $6.47 \times 10^{-12}$ | $-1.84 \times 10^{-12}$ | 95 |
| Terfenol-D | | 11.2 | -7.3 | $125 \times 10^{-12}$ | $-37.5 \times 10^{-12}$ | 10 |

Table 2 : Material properties for PZ27 (cited from Ferroperm[26]), Pic255 (cited from Physik Intrumente[27]), NiCoZn Ferrite (x=0.125) , and Terfenol-D.

**FIGURE CAPTIONS**

Fig 1. Sketches of the transversal coupling (top) and longitudinal coupling (bottom). The 1-2 plane coincides with the bases of the cylindrical ME sample.

Fig 2. Radial demagnetizing factor (triangle symbols) and axial demagnetizing factor (circle symbols) as a function of the thickness/diameter ratio of the magnetic material. Open circles: $\chi = 10000$. Filled circles: $\chi = 10$. Open triangles: $\chi = \infty$. Filled triangles: $\chi = 9$. Data are from D. X. Chen *et al*[16,17].

Fig 3. Transversal magnetoelectric coefficient $\alpha_{31}$ for the various Zn contents in ferrites. Top: slow-cooled ferrites. Bottom: quenched ferrites. All ferrites are pasted on PZ27 discs.

Fig 4. Transversal (dashed line) and longitudinal (dotted line) magnetoelectric coefficients for the sample # 0.5SC2/1. The solid line is the longitudinal ME coefficient calculated from the transversal one.

Fig 5. Magnetization and magnetization derivative measured for the slow-cooled ferrite with x=0.5 Zn content.

Fig 6. Transversal magnetoelectric coefficients for bilayers and trilayer made with NiCoZn ferrite (x=0.125) and PZ27. Thick lines correspond to experimental ME measurements. Thick solid line: 2mm/1mm ferrite/PZT bilayer (reference sample). Thick dashed line: 1mm/0.5mm ferrite/PZT bilayer. Thick dotted line: 1mm/1mm/1mm PZT/ferrite/PZT trilayer. The thin lines are curves deduced from the experimental one of the reference sample where demagnetizing effects were corrected. Thin dashed line: 1mm/0.5mm ferrite/PZT bilayer. Thin dotted line: 1mm/1mm/1mm PZT/ferrite/PZT trilayer.

Fig 7. Transversal magnetoelectric coefficients for $(Ni_{0.973}\ Co_{0.027})_{0.875}Zn_{0.125}Fe_2O_4$/PZ27 bilayers and trilayer.. Solid line: 1mm/2mm PZT/ferrite bilayer (reference sample). Dotted-dashed line: 0.5mm/2mm/0.5mm PZT/ferrite/PZT trilayer. Dotted line: 1mm/1mm/1mm PZT/ferrite/PZT trilayer. Dashed line: 0.5mm/0.5mm/0.5mm ferrite/PZT/ferrite trilayer. All the lines correspond to experimental ME measurements.

Fig 8: Strain curves versus applied external field along the parallel direction for the ME sample #0.125SC2/1 (reference sample). The closed circles correspond to strain measurement on the ferrite face, and open circles correspond to strain measurement on the PZT face. The magnetic field and strain measurements are parallel (and in the disc plane).

Fig 9. Top view of the ME device when circumferential AC magnetic field is forced within the ferrite. The PZT top electrode is split into two half electrodes (in grey) which produce two opposite voltages *V* and *V'*. Only AC voltage produced by a half piezoelectric layer is measured.

Fig 10. Transversal magnetoelectric coefficients when AC magnetic fields are forced within the ferrite. Solid line: sample # 0.25Q2/1 (quenched ferrite). Dashed line: sample # 0.25SC2/1 (slow cooled ferrite).

Fig 11. Transversal magnetoelectric coefficient $\alpha_{31}$ for sample # 0.125SC2/1 (solid line) and Terfenol-D/Pz27 2mm/1mm bilayer (dotted line).

Fig 12. Magnetostriction curves versus applied external field along parallel (closed circles) and perpendicular (open circles) directions for a Terfenol-D disc. The magnetic field and strain measurements are in the disc plane.

Fig 13. Magnetostriction curves versus applied external field along parallel (closed circles) and perpendicular (open circles) directions for a ferrite disc with x=0.125. The magnetic field and strain measurements are in the disc plane.

Fig 14. Black line: voltage obtains with an commercial active current probe (gain: 100mV/A). Grey line: ME voltage obtains at the electrical output of the piezoelectric material. The current have a square waveform at 1kHz and 10A peak.

Fig 15. Black line: voltage obtains with an commercial active current probe (gain: 100mV/A). Grey line: ME voltage obtains at the electrical output of the piezoelectric material. The current have a triangle waveform at 1kHz and 10A peak.

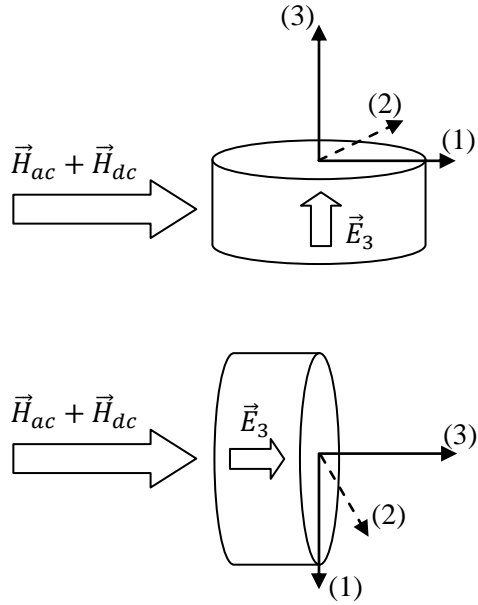

Fig 1. Sketches of the transversal coupling (top) and longitudinal coupling (bottom). The 1-2 plane coincides with the bases of the cylindrical ME sample.

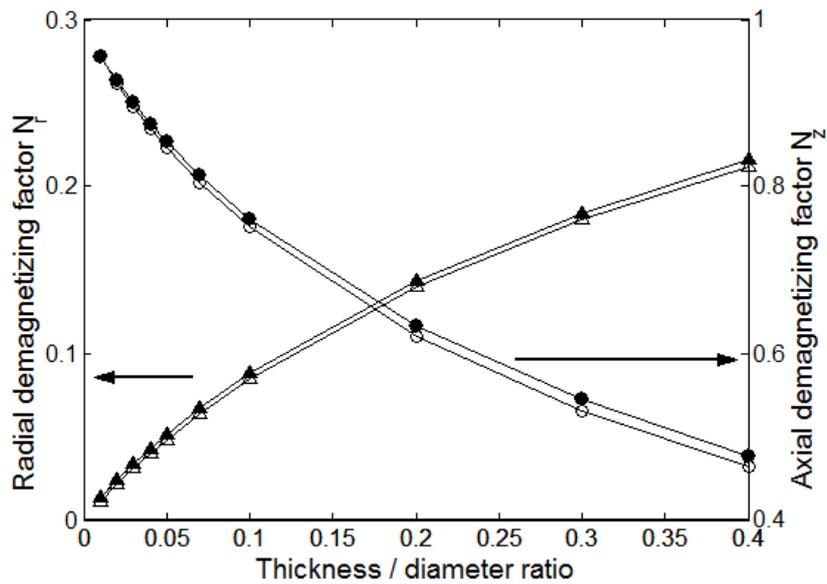

Fig 2. Radial demagnetizing factor (triangle symbols) and axial demagnetizing factor (circle symbols) as a function of the thickness/diameter ratio of the magnetic material. Open circles: $\chi = 10000$. Filled circles: $\chi = 10$. Open triangles: $\chi = \infty$. Filled triangles: $\chi = 9$. Data are from D. X. Chen *et al*[16,17].

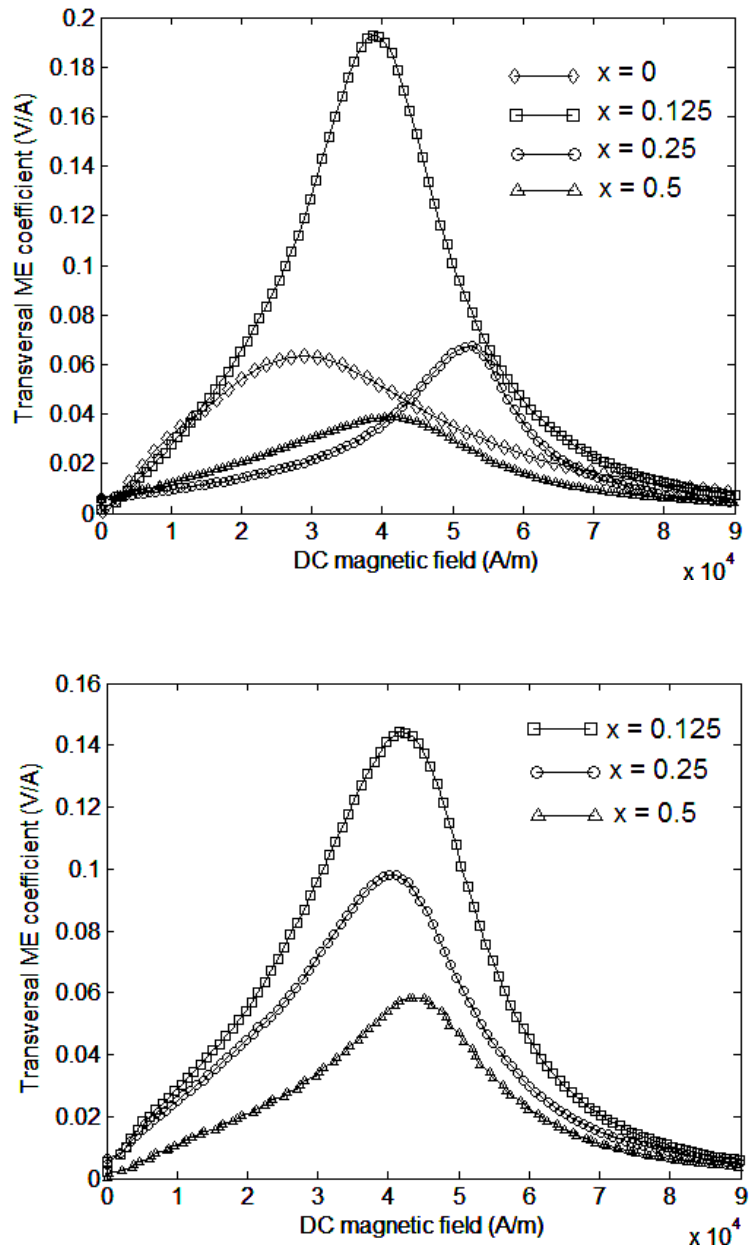

Fig 3. Transversal magnetoelectric coefficient $\alpha_{31}$ for the various Zn contents in ferrites. Top: slow-cooled ferrites. Bottom: quenched ferrites. All ferrites are pasted on PZ27 discs.

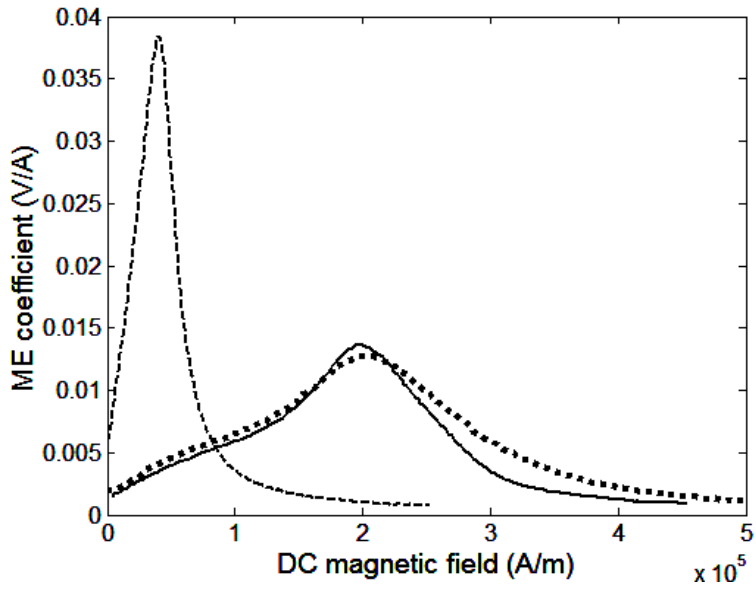

Fig 4. Transversal (dashed line) and longitudinal (dotted line) magnetoelectric coefficients for $(Ni_{0.973} Co_{0.027})_{0.5}Zn_{0.5}Fe_2O_4/Pz27$ bilayer. The solid line is the longitudinal ME coefficient calculated from the transversal one.

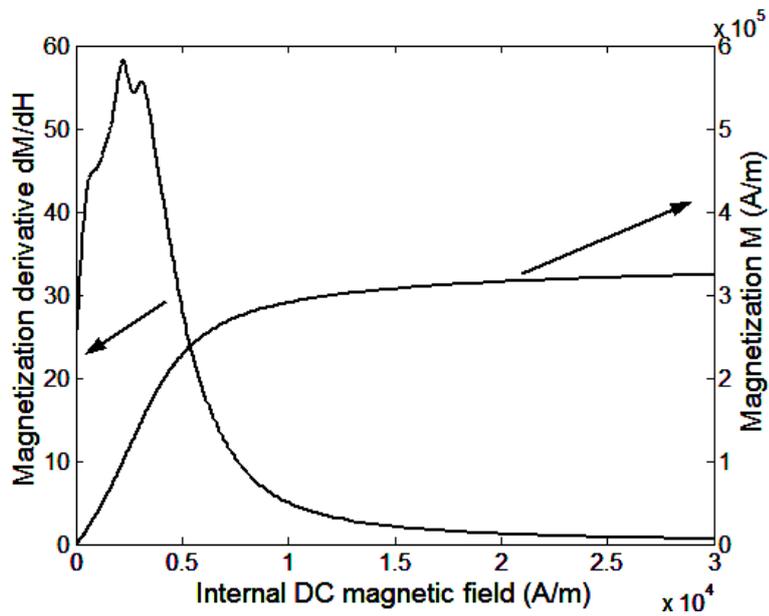

Fig 5. Magnetization and magnetization derivative measured for $(Ni_{0.973} Co_{0.027})_{0.5}Zn_{0.5}Fe_2O_4$ slow-cooled ferrite.

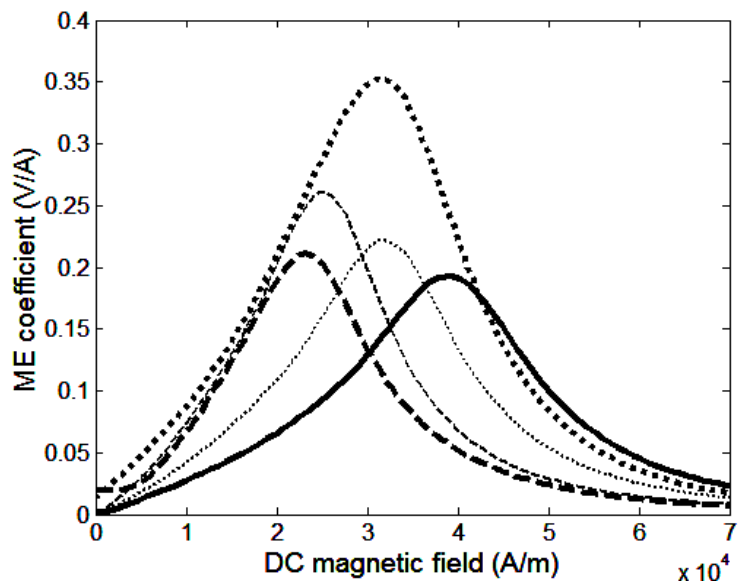

Fig 6. Transversal magnetoelectric coefficients for $(Ni_{0.973}\ Co_{0.027})_{0.875}Zn_{0.125}Fe_2O_4/PZ27$ bilayers and trilayer. Thick lines correspond to experimental ME measurements. Thick solid line: 1mm/2mm PZT/ferrite bilayer (reference sample). Thick dashed line: 0.5mm/1mm PZT/ferrite bilayer. Thick dotted line: 1mm/1mm/1mm PZT/ferrite/PZT trilayer. The thin lines are curves deduced from the experimental one of the reference sample where demagnetizing effects were corrected. Thin dashed line: 1mm/2mm PZT/ferrite bilayer. Thin dotted line: 1mm/1mm/1mm PZT/ferrite/PZT trilayer.

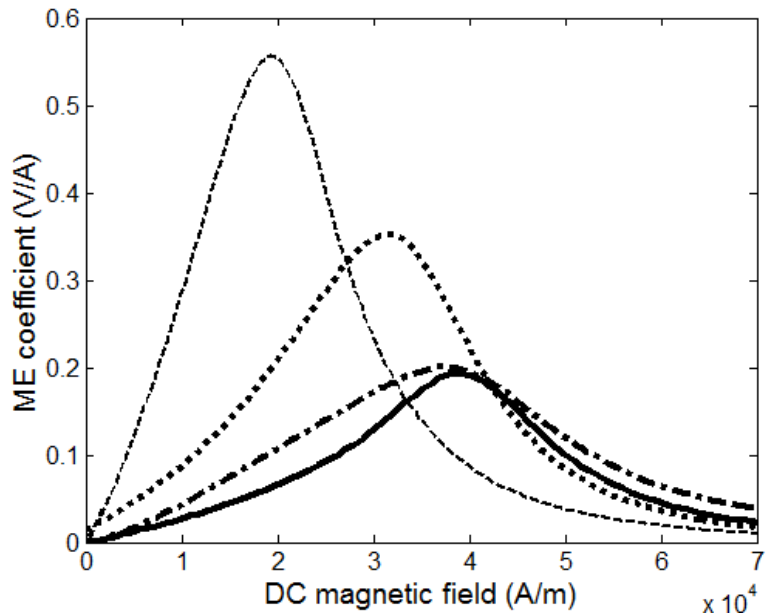

Fig 7. Transversal magnetoelectric coefficients for $(Ni_{0.973}Co_{0.027})_{0.875}Zn_{0.125}Fe_2O_4$/PZ27 bilayers and trilayer.. Solid line: 1mm/2mm PZT/ferrite bilayer (reference sample). Dotted-dashed line: 0.5mm/2mm/0.5mm PZT/ferrite/PZT trilayer. Dotted line: 1mm/1mm/1mm PZT/ferrite/PZT trilayer. Dashed line: 0.5mm/0.5mm/0.5mm ferrite/PZT/ferrite trilayer. All the lines correspond to experimental ME measurements.

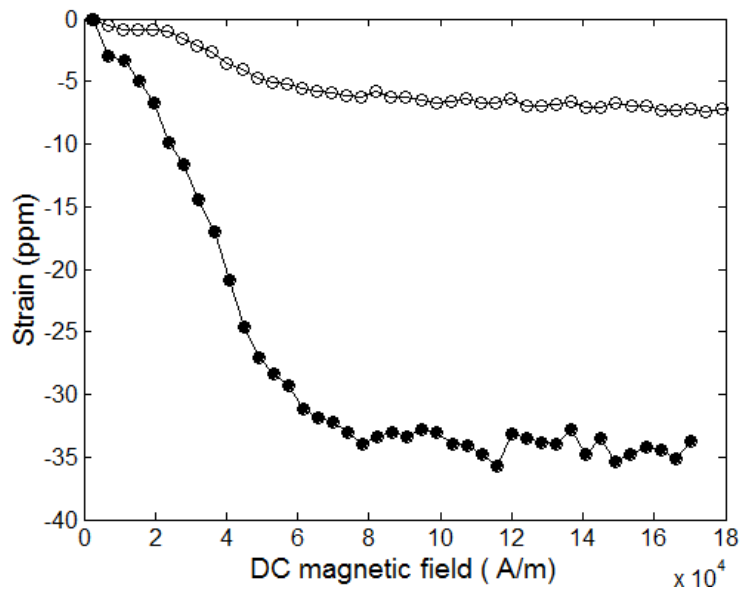

Fig 8: Strain curves versus applied external field along the parallel direction for the ME sample #0.125SC2/1 (reference sample). The closed circles correspond to strain measurement on the ferrite face, and open circles correspond to strain measurement on the PZT face. The magnetic field and strain measurements are parallel (and in the disc plane).

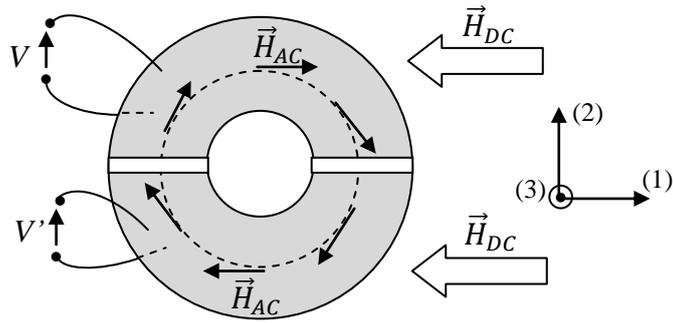

Fig 9: Top view of the ME device when circumferential AC magnetic field is forced within the ferrite. The PZT top electrode is split into two half electrodes (in grey) which produce two opposite voltages $V$ and $V'$. Only AC voltage produced by a half piezoelectric layer is measured.

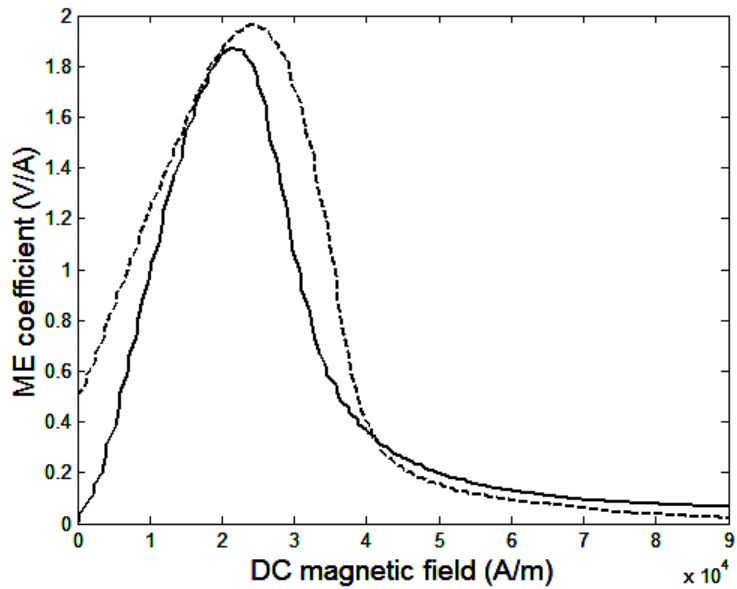

Fig 10. Transversal magnetoelectric coefficients for $(Ni_{0.973}\ Co_{0.027})_{0.75}Zn_{0.25}Fe_2O_4$/Pic255 bilayers when AC magnetic fields are forced within the ferrite. Solid line: quenched ferrite. Dashed line: slow cooled ferrite.

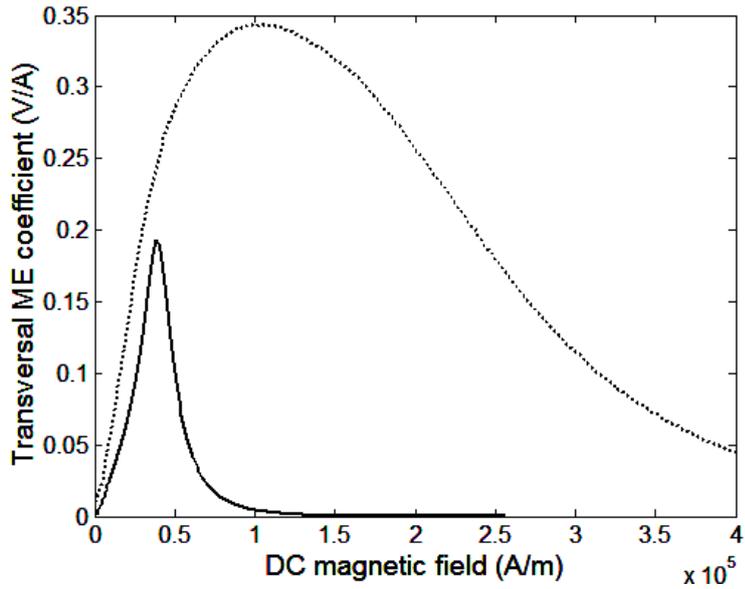

Fig 11. Transversal magnetoelectric coefficient $\alpha_{31}$ for $(Ni_{0.973}Co_{0.027})_{0.875}Zn_{0.125}Fe_2O_4$/Pz27 2mm/ 1mm bilayer (solid line) and Terfenol-D/Pz27 2mm/1mm bilayer (dotted line).

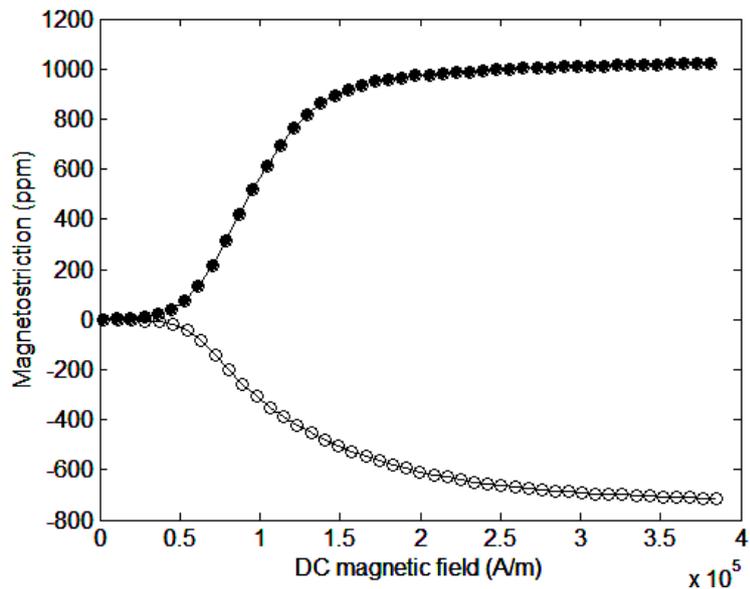

Fig 12. Magnetostriction curves versus applied external field along parallel (closed circles) and perpendicular (open circles) directions for a Terfenol-D disc. The magnetic field and strain measurements are in the disc plane.

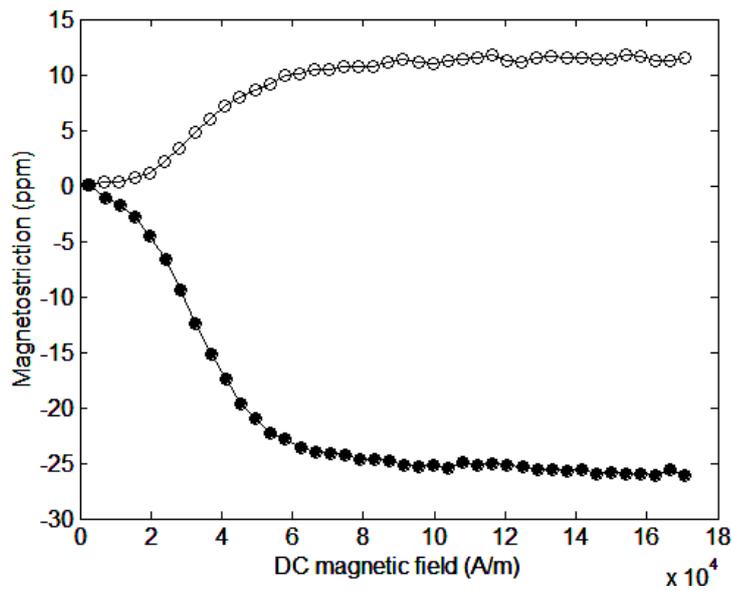

Fig 13. Magnetostriction curves versus applied external field along parallel (closed circles) and perpendicular (open circles) directions for a ferrite disc with x=0.125. The magnetic field and strain measurements are in the disc plane.

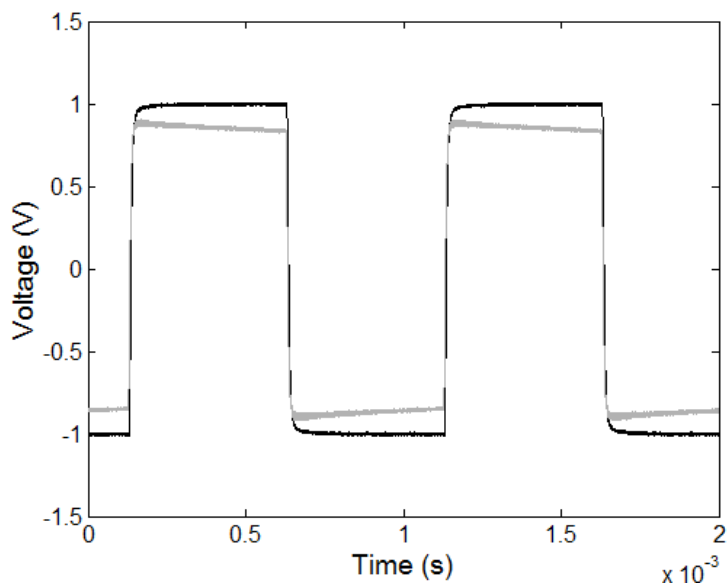

Fig 14. Black line: voltage obtains with an commercial active current probe (gain: 100mV/A). Grey line: ME voltage obtains at the electrical output of the piezoelectric material. The current have a square waveform at 1kHz and 10A peak.

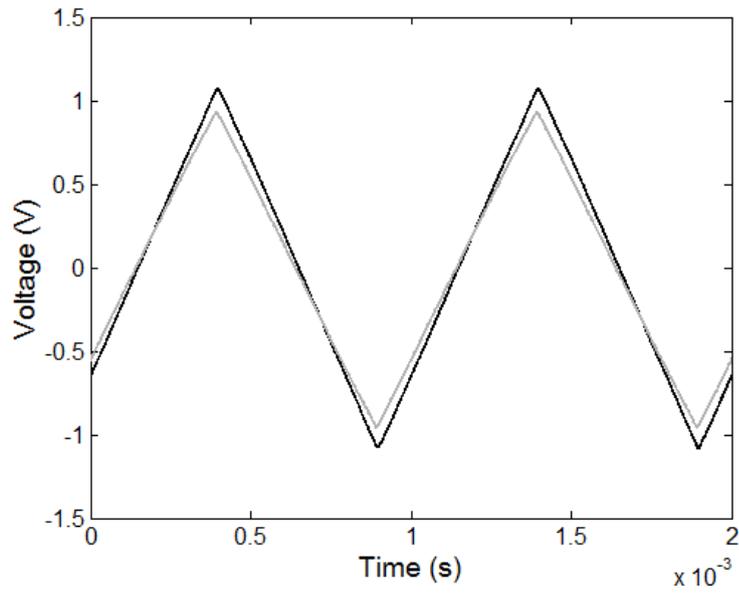

Fig 15. Black line: voltage obtains with an commercial active current probe (gain: 100mV/A). Grey line: ME voltage obtains at the electrical output of the piezoelectric material. The current have a triangle waveform at 1kHz and 10A peak.